\pdfoutput=1
\documentclass[%
 notitlepage,amsmath,amssymb,
 aps,
]{revtex4-1}
\usepackage[pdftex]{graphicx}
\usepackage{bm}
\usepackage{hyperref}
\usepackage{amsmath}
\usepackage[pdftex]{color}
\usepackage{subfigure}

\newcommand{\textunderscript}[1]{$_{\text{#1}}$}
\renewcommand{\figurename}{\textbf{Figure}}
\begin{document}

\title{Two types of all-optical magnetization switching mechanisms using \\ femtosecond laser pulses}

\author{M.S. El Hadri}
\affiliation {Institut Jean Lamour, UMR CNRS 7198, Universit\'{e} de Lorraine,  BP 70239, F-54506, Vandoeuvre-l\`{e}s-Nancy, France}

\author{P. Pirro}
\affiliation {Institut Jean Lamour, UMR CNRS 7198, Universit\'{e} de Lorraine,  BP 70239, F-54506, Vandoeuvre-l\`{e}s-Nancy, France}

\author{C.-H. Lambert}
\affiliation {Institut Jean Lamour, UMR CNRS 7198, Universit\'{e} de Lorraine,  BP 70239, F-54506, Vandoeuvre-l\`{e}s-Nancy, France}

\author{S. Petit-Watelot}
\affiliation {Institut Jean Lamour, UMR CNRS 7198, Universit\'{e} de Lorraine,  BP 70239, F-54506, Vandoeuvre-l\`{e}s-Nancy, France}

\author{Y. Quessab}
\affiliation {Institut Jean Lamour, UMR CNRS 7198, Universit\'{e} de Lorraine,  BP 70239, F-54506, Vandoeuvre-l\`{e}s-Nancy, France}

\author{M. Hehn}
\affiliation {Institut Jean Lamour, UMR CNRS 7198, Universit\'{e} de Lorraine,  BP 70239, F-54506, Vandoeuvre-l\`{e}s-Nancy, France}

\author{F. Montaigne}
\affiliation {Institut Jean Lamour, UMR CNRS 7198, Universit\'{e} de Lorraine,  BP 70239, F-54506, Vandoeuvre-l\`{e}s-Nancy, France}

\author{G. Malinowski}
\affiliation {Institut Jean Lamour, UMR CNRS 7198, Universit\'{e} de Lorraine,  BP 70239, F-54506, Vandoeuvre-l\`{e}s-Nancy, France}

\author{S. Mangin}
\affiliation {Institut Jean Lamour, UMR CNRS 7198, Universit\'{e} de Lorraine,  BP 70239, F-54506, Vandoeuvre-l\`{e}s-Nancy, France}

\date{\today}

\begin{abstract}
Magnetization manipulation in the absence of an external magnetic field is a topic of great interest, since many novel physical phenomena need to be understood and promising new applications can be imagined. Cutting-edge experiments have shown the capability to switch the magnetization of magnetic thin films using ultrashort polarized laser pulses. In 2007, it was first observed that the magnetization switching for GdFeCo alloy thin films was helicity-dependent and later helicity-independent switching was also demonstrated on the same material. Recently, all-optical switching has also been discovered for a much larger variety of magnetic materials (ferrimagnetic, ferromagnetic films and granular nanostructures), where the theoretical models explaining the switching in GdFeCo films do not appear to apply, thus questioning the uniqueness of the microscopic origin of all-optical switching. Here, we show that two different all-optical switching mechanisms can be distinguished; a ``single pulse" switching and a ``cumulative" switching process whose rich microscopic origin is discussed. We demonstrate that the latter is a two-step mechanism; a heat-driven demagnetization followed by a helicity-dependent remagnetization. This is achieved by an all-electrical and time-dependent investigation of the all-optical switching in ferrimagnetic and ferromagnetic Hall crosses via the anomalous Hall effect, enabling to probe the all-optical switching on different timescales.

\end{abstract}
\maketitle

The control of magnetism without magnetic fields is an emergent field of research due to the prospect of understanding novel physical mechanisms and impacting application fields such as low-power electronics and magnetic data storage. Promising mechanisms of magnetization manipulation include the tuning of magnetic properties using an electric field \cite{Ohno2000,Chiba2003,Chiba2008}, as well as the spin transfer torque (STT) switching arising from several mechanisms such as the direct momentum transfer \cite{Slonczewski1996,Berger1996,Katine2000}, the spin-orbit interaction \cite{Liu2012,Liu2012b} or the spin-Seebeck effect \cite{Hatami2007,Uchida2008,Yu2010}. Nevertheless, an intriguing new possibility to switch the magnetization consists of using ultrashort laser pulses without any other stimuli. Indeed, recent experiments have shown the possibility to perform magnetization reversal of ferrimagnetic GdFeCo alloy films on a picosecond (ps) timescale using only femtosecond laser pulses \cite{Stanciu2007,Kirilyuk2010}. Thus, this all-optical switching  is of great interest and has the potential to be implemented in spintronic devices like STT based memories \cite{Kent2015,Parkin2015} or heat-assisted magnetic recording (HAMR) \cite{Kryder2008,Stipe2010}. In this context, the discovery of all-optical switching (AOS) in a variety of ferrimagnetic materials, including ferrimagnetic alloys with various rare earth (RE) \cite{Alebrand2012}, ferrimagnetic multilayers and heterostructures \cite{Mangin2014,Schubert2014}, as well as RE-free synthetic ferrimagnetic heterostructures has attracted large interest. Very recently, all-optical switching has even been demonstrated in pure ferromagnetic films such as [Pt/Co] and [Ni/Co] multilayers and FePt granular media \cite{Lambert2014}.

The large variety of materials where AOS could be observed immediately raised the question of the uniqueness of its microscopic origin. Indeed, for the initially investigated GdFeCo alloy films, both the experiments and the simulations demonstrated in literature agree on the microscopic origin of AOS, which is an ultrafast and pure thermal effect \cite{Vahaplar2012,Ostler2012,Kirilyuk2012}. In this particular ferrimagnetic material, the net magnetization of the transition metal (TM) and rare earth (RE) sublattices are antiferromagnetically exchange coupled and the spin dynamics of the two sublattices are different since the TM component demagnetizes faster than the RE one. Considering those two properties a measured transient ferromagnetic-like state was explained to be behind the AOS effect \cite{Radu2011}. However, this mechanism can obviously not be applied to the ferromagnetic films and grains, which do not present an antiferromagnetic exchange coupling between two sublattices. In addition, all-optical switching of GdFeCo alloy films is found to be independent of the circular helicity of laser pulses with sufficiently high fluence, which is explained by heat pulses only \cite{Radu2011}. In contrast, all-optical switching in the other ferrimagnetic and ferromagnetic materials appears to be always helicity-dependent \cite{Alebrand2012b,Alebrand2014,Hassdenteufel2015,El-Hadri2016}.
 
To investigate the microscopic mechanisms involved in the magnetization switching of different materials, we have combined two approaches. The first approach consists of imaging the magnetic contrast after one or several pulses, while the second approach is based on a time-dependent electrical method by measuring the magnetization switching in Hall crosses via the anomalous Hall effect (AHE) \cite{Luttinger1958,Berger1970}. The latter approach enables to probe the all-optical switching over several time scales as well as to quantify the switching ratio for different laser parameters. It bridges the time gap between the ultrafast and the quasi-static optical methods to investigate magnetization switching. In the following, we will show experimental evidence that two different all-optical switching mechanisms exist; a single-shot switching as previously demonstrated in GdFeCo films, and a cumulative multi-pulse switching for Co/Pt and TbCo films. The rich microscopic origin is thereafter discussed. \\

\noindent
\textbf{Results} 
\\
\textbf{Outline of the experiments and samples structure.} To explore the mechanism of all-optical switching, we combined magneto-optical imaging and magneto-transport measurements of multiple femtosecond (fs) laser pulse switching of ferri- and ferro-magnetic materials. In the magneto-transport measurements, the magnetic response is probed via the AHE with a configuration as displayed in Fig. 1a. The AOS process has often been studied on the ultrashort timescale via time-resolved techniques, including pump-probe technique with measurement duration up to the nanosecond (ns) scale, or on a quasi-static time scale via magneto-optical Kerr/Faraday imaging \cite{Stanciu2007,Mangin2014}.  Thus, the main feature of using an electrical probe via AHE is the possibility to measure the magnetization change over an intermediate and wide timescale ranging from 1 $\mu$s to a few seconds. As it has recently been shown that all-optical switching in [Pt/Co] multilayers occurs only in a rim at the edge of a demagnetized area \cite{Lambert2014}, the position of the fs laser beam in such experiment is maintained fixed and off-centered so that the rim where AOS takes place overlaps with the central area of the Hall cross as shown in Fig. 1b.

\begin{figure}[h]
\begin{center}
\scalebox{1}{\includegraphics[width=14 cm, clip]{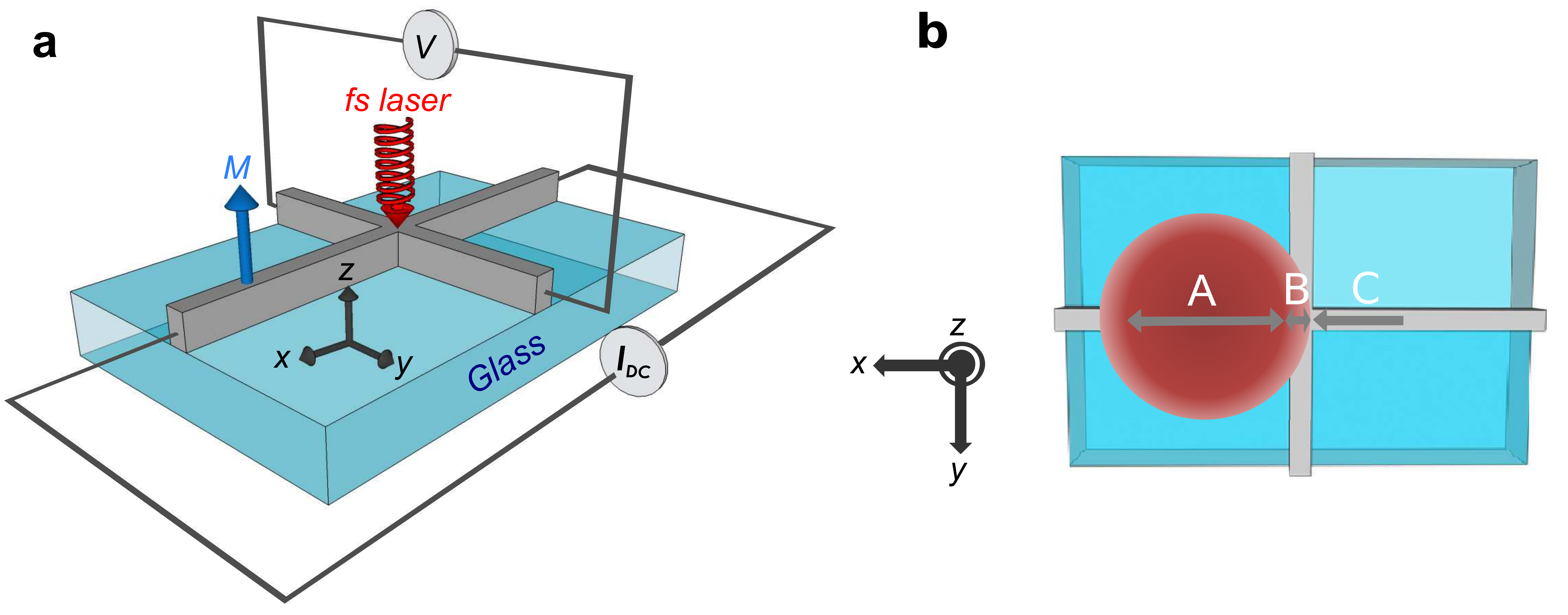}}
\end{center}
\makeatletter 
\renewcommand{\thefigure}{\textbf{1}}
\caption{\label{sample} \textbf{Experimental set-up schematic of fs laser beam switching of Hall crosses.}  (\textbf{a}) Schematic of the studied 5-$\mu$m-wide Hall crosses with perpendicular magnetization (\textit{z} axis); a DC current is injected along the \textit{x} direction while the anomalous Hall voltage \textit{$V_{Hall}$} is measured along the \textit{y} direction with a temporal resolution of 1 $\mu$s. A fs laser beam with a FWHM of 60 $\mu$m illuminates the film plane along the z direction, at a fixed position about 40 $\mu$m from the center of the Hall cross along x direction. (\textbf{b}) Schematic representation of the different areas of the fs laser beam: ``A" where multiple magnetic domains are obtained, ``B" where AOS is obtained and ``C" where no change of magnetization is induced.} 
\end{figure}

The magneto-transport measurements are combined with optical studies of the response of the ferri-and ferromagnetic continuous films exposed to a single fs pulse using static Faraday microscopy, in order to image the magnetic domains and verify the consistency of the all-electrical measurement of AOS performed on Hall crosses.  For both experiments presented here, three different samples were investigated under the same conditions, namely two ferrimagnetic alloys: Tb\textunderscript{27}Co\textunderscript{73}(20 nm) and Gd\textunderscript{28}Fe\textunderscript{48}Co\textunderscript{24}(20 nm) which were capped with a Pt layer and a Ta layer respectively to prevent sample oxidation; and one ferromagnetic multilayer Pt(4.5 nm)/Co(0.6 nm)/Pt(4.5 nm), in which the Pt layers induce a strong perpendicular magnetic anisotropy and the top Pt layer also prevents sample oxidation. All of these three films were grown by DC sputtering on a Glass/Ta(3 nm) substrate, and show perpendicular magnetization in remanence. \\

\noindent
\textbf{Single-pulse switching of GdFeCo continuous films and Hall crosses.} 
We initially experimentally verified the helicity-independent switching of the ferrimagnetic Gd\textunderscript{28}Fe\textunderscript{48}Co\textunderscript{24} (20 nm) alloy film. Due to the zero net orbital momentum of Gd, the spin-orbit coupling of Gd-based alloys is supposed to be small, thus resulting in low magneto-crystalline anisotropy. Co is then used to induce perpendicular magnetic anisotropy. This ferrimagnetic sample shows magnetic compensation temperature (\textit{T\textunderscript{comp}}), at which the two collinear sublattice magnetizations \textit{M\textunderscript{Gd}} and \textit{M\textunderscript{FeCo}} compensate.
\begin{figure}
\makeatletter 
\renewcommand{\thefigure}{\textbf{2}}
    \centering
    \begin{subfigure} {}
       \includegraphics[width= 8.1 cm]{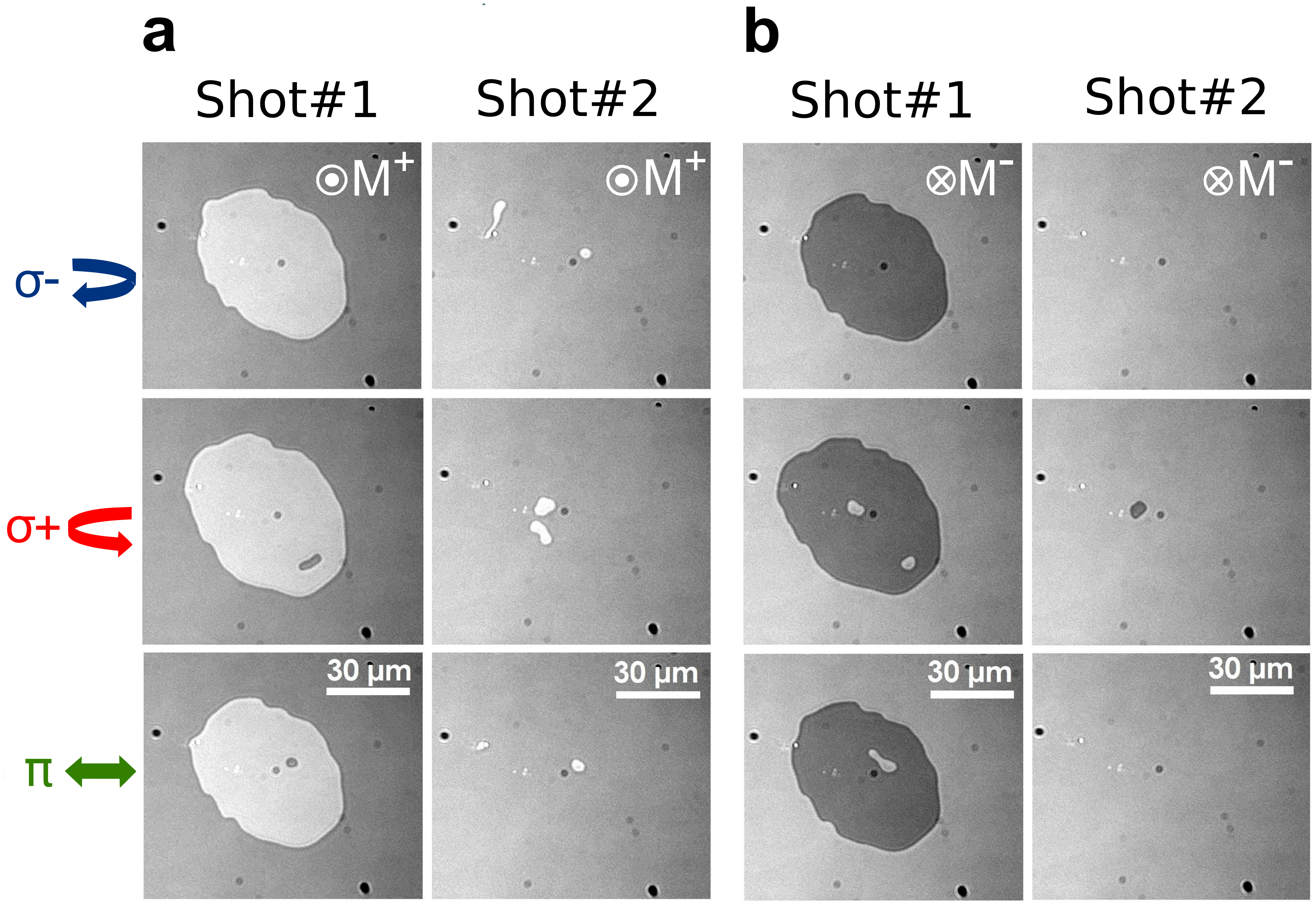}
    \end{subfigure}
    ~
    \begin{subfigure}  {}
        \includegraphics[width=7.6 cm]{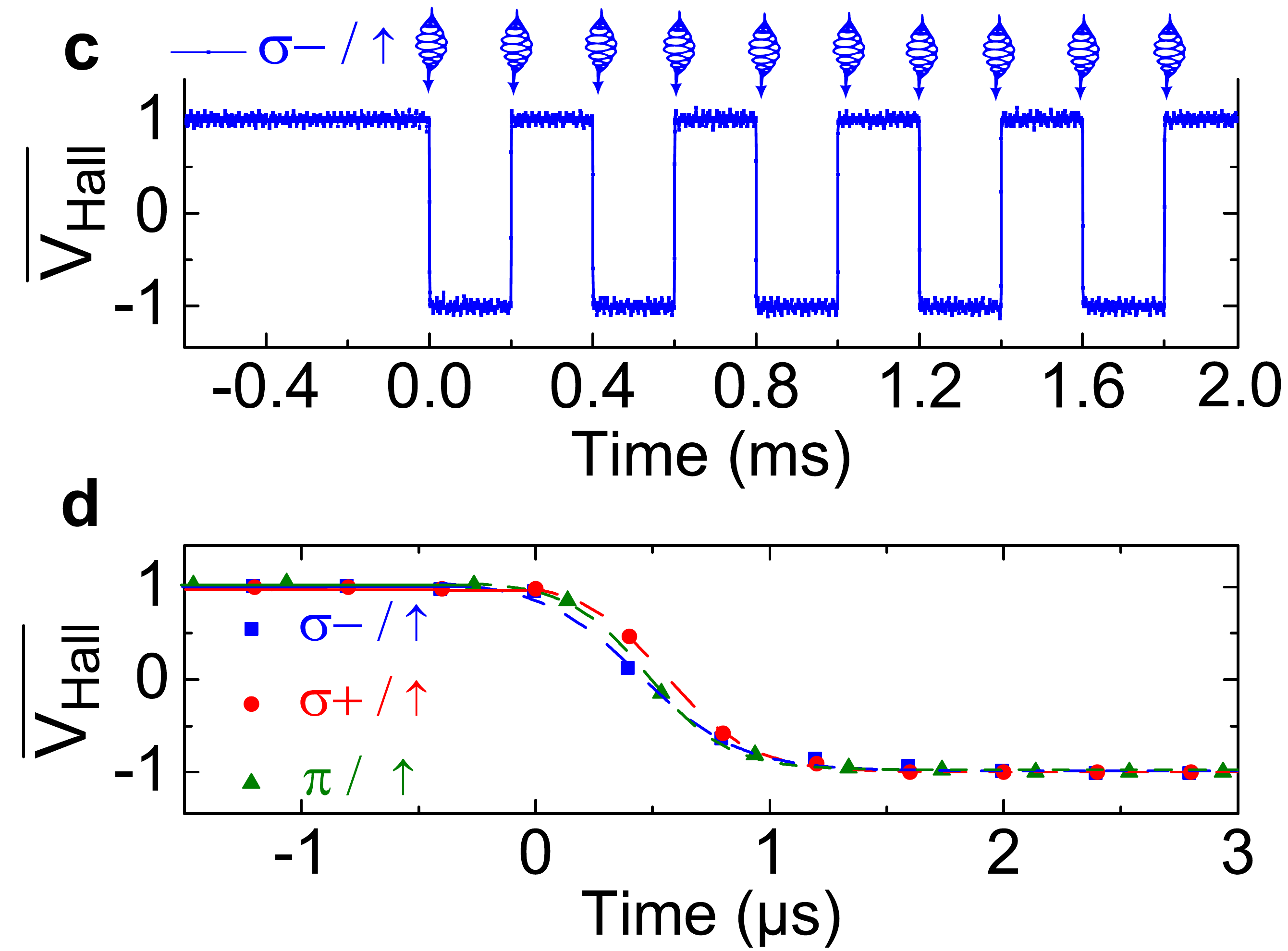}
    \end{subfigure}
    \caption{\label{sample} 
\textbf{Single shot and helicity-independent switching of Gd\textunderscript{28}Fe\textunderscript{48}Co\textunderscript{24} continuous film and patterned Hall cross.} (\textbf{a}) and (\textbf{b}) Magneto-optical Faraday images of a Gd\textunderscript{28}Fe\textunderscript{48}Co\textunderscript{24} continuous film, with an initial magnetization saturation ``up" and ``down", illuminated with two consecutive pulses with three different polarizations: from top to bottom, left circularly polarized pulse ($\sigma$-), right circularly polarized pulse ($\sigma$+) and linearly polarized pulse ($\pi$) with a fluence of 19 mJ/cm$^2$. Each of the two laser pulses illuminates the same region of the continuous film and induces a total magnetization reversal. The white (resp. dark) contrast in (\textbf{a}) (resp. (\textbf{b})) corresponds to a reversal to down (resp. to up). (\textbf{c}) Electrical measurement of magnetization reversal of the patterned Gd\textunderscript{28}Fe\textunderscript{48}Co\textunderscript{24} Hall cross under the action of ten consecutives pulses as marked with the blue pulses in the upper row, with initial saturation up with $\sigma$- polarization and a repetition rate of 5 kHz. The Hall voltage signal is normalized to 1 and -1 corresponding to the positive and negative magnetization saturations. Each of the ten laser pulses illuminates the same region of the Hall cross and reverses the magnetization within it. (\textbf{d}) Reversal of the anomalous Hall voltage \textit{$V_{Hall}$} in the patterned Gd\textunderscript{28}Fe\textunderscript{48}Co\textunderscript{24} Hall cross under the action of a single laser pulse with three different polarizations and a fluence of 19 mJ/cm$^2$. The reversal time of the magnetization is shorter than the temporal resolution of the measurement limited to 1 $\mu$s. Dashed lines are guides to the eyes.}
\end{figure}

We study the response of the  Gd\textunderscript{28}Fe\textunderscript{48}Co\textunderscript{24} continuous film to the action of 35-fs single pulses with photon energy of 1.55 eV corresponding to a wavelength of 800 nm for three different polarizations, the two circular polarizations and the linear one. Note that the magneto-optical Faraday microscopy is mainly sensitive to the perpendicular component of the FeCo sublattice magnetization. The studied continuous film is initially saturated up or down, prior to being excited with two consecutive pulses with the three different polarizations, as shown in Fig. 2a and Fig. 2b, respectively. Independently of the initial saturation and the polarization of the pulse, one can see that a total reversal of magnetization occurs in the irradiated area after the first pulse, whereas the second pulse switches magnetization back to the saturated state. We now implement this helicity-independent switching into a GdFeCo based Hall cross. As displayed in Fig. 2c,d, the magnetization of the Gd\textunderscript{28}Fe\textunderscript{48}Co\textunderscript{24} Hall cross is totally reversed after the action of a single pulse, for each pulse in a set of ten pulses. The same behavior has been found for all six possible combinations of initial saturation and light polarization. From Fig. 2d, the reversal time of the anomalous Hall voltage coincides with the temporal resolution of the measurement that equals 1 $\mu$s, thus indicating that the magnetization reversal actually takes a shorter time. These findings are in agreement with previous results shown for similar GdFeCo alloys \cite{Ostler2012}. Identical results were obtained by illuminating the Gd\textunderscript{28}Fe\textunderscript{48}Co\textunderscript{24} Hall cross with a set of 1000 consecutives pulses. \\

\begin{figure}[h]
\begin{center}
\scalebox{1}{\includegraphics[width=8 cm, clip]{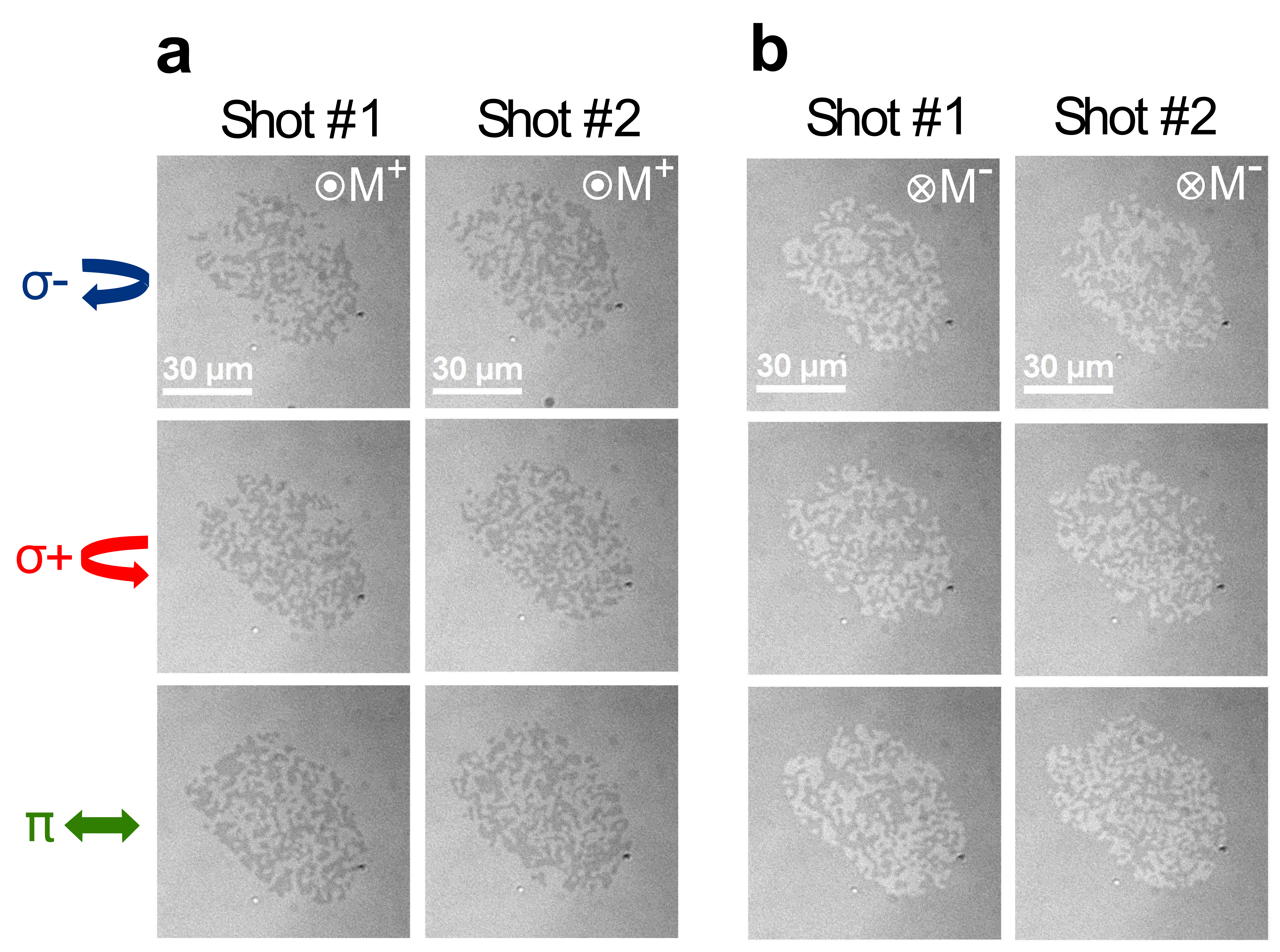}}
\end{center}
\makeatletter 
\renewcommand{\thefigure}{\textbf{3}}
\caption{\label{sample} \textbf{Single shot induced thermal demagnetization of Pt/Co/Pt continuous films.} (\textbf{a}) and (\textbf{b}) Magneto-optical Faraday images of a Pt(4.5 nm)/Co(0.6 nm)/Pt(4.5 nm) continuous film, with initial magnetization saturation ``up" and ``down", illuminated with two consecutive pulses with three different polarizations and with an energy per pulse of 14 mJ/cm$^2$. Each of the two laser pulses induces thermal demagnetization of the irradiated area. The dark (resp. white) contrast in (\textbf{a}) (resp. (\textbf{b})) corresponds to a reversal to down (resp. to up).}
\end{figure}

\noindent
\textbf{Single-pulse induced thermal demagnetization of Pt/Co/Pt continuous films.} As already mentioned, the recently demonstrated all-optical switching in [Pt/Co] multilayers using a fs laser beam is helicity-dependent for a laser fluence ranging from the threshold below which the laser does not affect the magnetization to the damage threshold, and occurs only in a rim at the edge of a demagnetized area \cite{Lambert2014}. This demagnetized area is most likely induced by laser heating in the vicinity of the ordering temperature (e.g. Curie temperature \textit{T\textunderscript{c}}), and is characterized by the formation of magnetic domains with random orientations up and down. To give novel insight into the switching mechanism in these materials, we now study the response of a Pt/Co/Pt continuous film to the action of two consecutive 35-fs pulses, thus the analog experiment to GdFeCo films shown in Fig. 2a. According to the Faraday images in Fig. 3a,b, each of these two pulses induces a random thermal demagnetization of the irradiated area, independently of the initial saturation and the polarization of the pulse. The domains distribution in the demagnetized area was quantified, leading to two close average values for magnetic domains up and down of approximately 50 $\pm$ 2$\%$. Moreover, these measurements show clearly that the expected rim of helicity-dependent switching does not emerge after the two consecutives pulses and for any of the six combinations of polarization and initial saturation (see Supplementary Fig. 1). Indeed, the fluence used in this experiment is high enough to induce a helicity-dependent switching with a laser spot swept over the sample, corresponding to an exposure time of the order of one second. More importantly, only random thermal demagnetization is measured by increasing the laser fluence up to the damage fluence threshold. \vspace{0.7cm} \\

\noindent
\textbf{Multiple-pulse helicity-dependent switching of Pt/Co/Pt Hall crosses.} To investigate the earlier reported cumulative as well as pronounced helicity-dependent switching process in such ferromagnetic material, we now continue by electrically quantifying the magnetization change in a ferromagnetic Pt/Co/Pt based Hall cross under the illumination of a set of consecutive pulses. Figure 4 shows the time-dependent evolution of the anomalous Hall voltage \textit{$V_{Hall}$} of a 5-$\mu$m-wide  Pt(4.5 nm)/Co(0.6 nm)/Pt(4.5 nm) Hall cross under the action 600 consecutive 35-fs pulses for different polarizations and initial saturations. To avoid damaging the studied Hall cross, this multiple shot measurement was performed with a laser fluence of 10 mJ/cm$^2$, which is significantly lower than the one used in the previous single shot experiment on the Pt/Co/Pt continuous film. One can see from Fig. 4 that this fluence range enables the decomposition of the switching of the anomalous Hall voltage into a two regimes process taking place at two different timescales; a helicity-independent drop induced by 6 pulses within the first 1 ms and a helicity-dependent reversal of the Hall voltage on a several ten ms timescale.

First, as shown in Fig. 4a,c, \textit{$V_{Hall}$} follows a step-like decrease within the first 1 ms due to the action of the first laser pulses. Each jump of \textit{$V_{Hall}$} is induced by only one pulse and occurs within the experimental temporal resolution of 1 $\mu$s, thus indicating that each jump is actually taking place on a shorter time scale. These findings are a hint that the magnetization of the Hall cross is subject to multiple pulse demagnetization. Note that this demagnetization is helicity-independent, thus highlighting the important role of heat in this process  \cite{Beaurepaire1996}. Moreover, as can be seen in Fig. 4a,c, \textit{$V_{Hall}$}  is slowly increasing toward its initial value after each excitation pulse, which might be due to a slow magnetization relaxation toward its thermal equilibrium induced by the cooling of the material. One may argue that the variation of the anomalous Hall voltage might be due to a change of the Hall resistivity induced by the temperature increase. Nevertheless, by adding an external magnetic field which saturates the laser-illuminated Hall cross after each pulse, no change of the anomalous Hall voltage is measured. This excludes a significant change of the Hall resistivity and is in agreement with previous studies (see Supplementary Fig. 2) \cite{El-Hadri2016}.

One of the most interesting findings here is that the signature of the helicity starts to appear only after crossing the zero Hall voltage. Indeed, Fig. 4b,d clearly demonstrate that the anomalous Hall voltage and thus the magnetization of the Hall cross begin to gradually remagnetize to up (resp. down) under the action of left-($\sigma$-) (resp. right-($\sigma$+)) circular polarized pulses independently of the initial magnetization saturation. Note that only 90$\%$ of the reversal is achieved. This might be due to the small width of the switching rim of the laser spot compared to the Hall cross area. Furthermore, the complete switching of the Hall cross using circular helicities takes place only after an exposure time of 30 ms, thus demonstrating that with this laser power, approximately 150 pulses are needed to obtain a reliable switching. On the other hand, under the excitation of linearly polarized pulses, \textit{$V_{Hall}$} oscillates around the zero value up to 120 ms, meaning that multidomain configuration is permanently obtained with linear polarization. Thus, these experiments demonstrate that the all-optical switching of ferromagnetic Pt/Co/Pt films is helicity dependent, cumulative and only achieved after a thermal demagnetization, and hence is drastically different from the GdFeCo switching mechanism. \\
\begin{figure}[h]
\begin{center}
\scalebox{1}{\includegraphics[width=13 cm, clip]{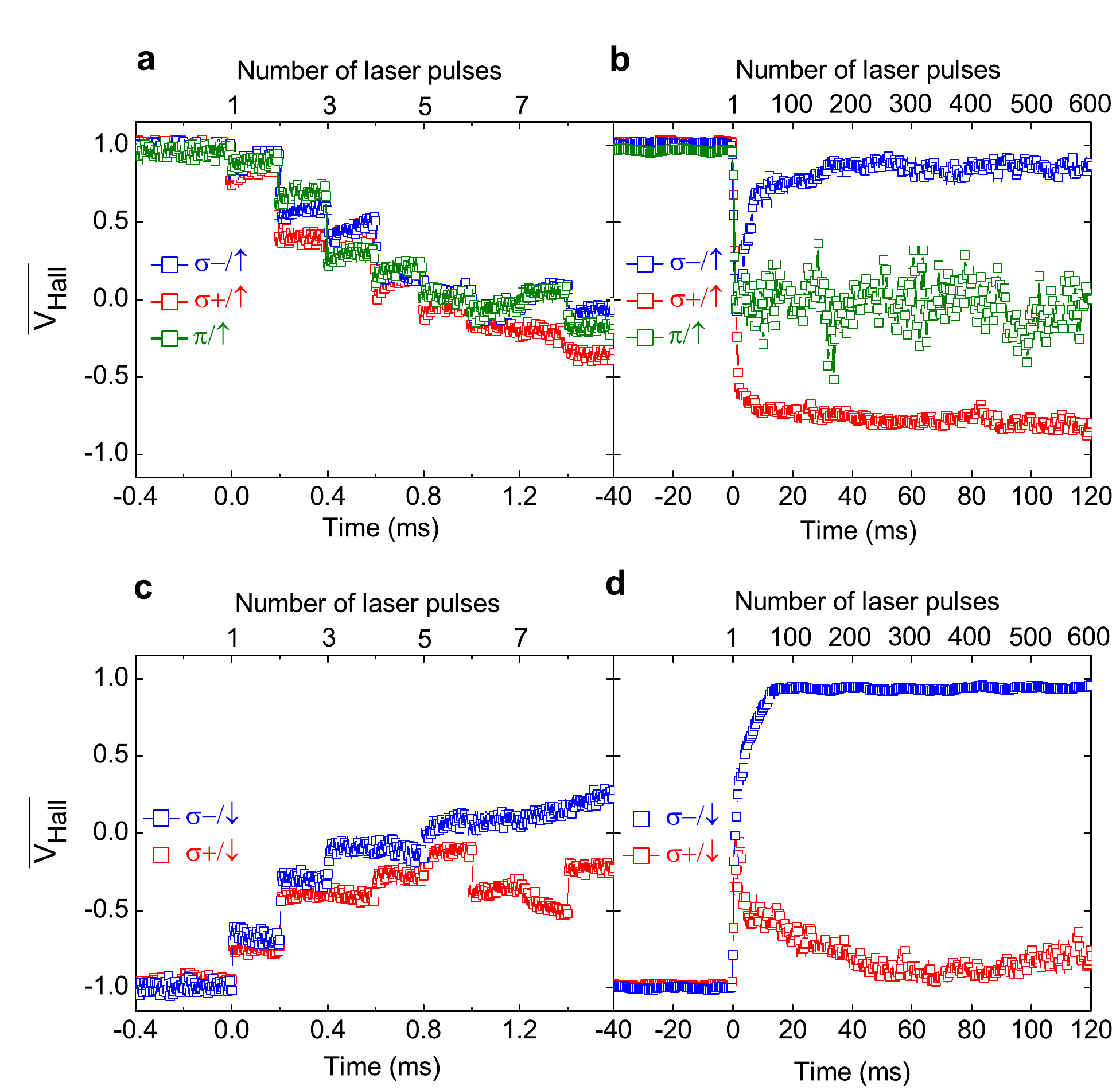}}
\end{center}
\makeatletter 
\renewcommand{\thefigure}{\textbf{4}}
\caption{\label{sample} \textbf{Multiple pulse helicity-dependent switching of Pt/Co/Pt Hall crosses.} (\textbf{a}) and (\textbf{b}) Time evolution of the anomalous Hall voltage of a 5-$\mu$m-wide Pt(4.5 nm)/Co(0.6 nm)/Pt(4.5 nm) Hall cross initially saturated ``up", under the action of a 35-fs laser beam with a 5 kHz repetition rate and a fluence of 10 mJ/cm$^2$. The corresponding number of laser pulses is shown in the upper row. The helicity-dependent switching of the studied Hall cross is governed by a two-step process at two different timescales: (\textbf{a}) helicity-independent demagnetization within the first 1ms. (\textbf{b}) helicity-dependent reversal measured on a 120 ms timescale. (\textbf{c}) and (\textbf{d}), the same behavior is measured as (\textbf{a}) and (\textbf{b}) with initial saturation of the Hall cross ``down". The experimental temporal resolution of the experiment is 1 $\mu$s. }
\end{figure}

\noindent
\textbf{Multiple-pulse helicity-dependent switching of TbCo Hall crosses.} It has been mentioned that previous studies have demonstrated that AOS is also apparent in a variety of ferrimagnetic materials such as TbCo alloys. Moreover, the multiple-shot switching in such ferrimagnetic films was obtained with a swept beam and was helicity-dependent from the switching threshold to the damage fluence threshold \cite{Alebrand2012,Mangin2014}. Counterintuitively, this behavior is at the opposite to the one measured on ferrimagnetic GdFeCo films for which the helicity-dependence was only obtained for a narrow window of fluence \cite{Khorsand2012}, and is similar to the one demonstrated on ferromagnetic [Pt/Co] films. This raises the questions of what are the major differences between the two ferrimagnetic TbCo and GdFeCo alloys and is the switching mechanism the same for TbCo alloys and ferromagnetic [Pt/Co] multilayers? The studied GdFeCo and TbCo alloy films show similar magnetic properties. Indeed, they have approximately the same RE concentration and both are dominated by the RE sublattice magnetization at room temperature. Moreover, they have a similar compensation temperature, which is above room temperature (\textit{T\textunderscript{comp}} = 500 K). On the other hand, at the opposite of the zero net orbital momentum of Gd, the orbital momentum in Tb is large, thus leading to a stronger spin-orbit coupling and larger magneto-crystalline anisotropy. It was also shown that Gd- and Tb-based alloys show different ultrafast spin dynamics under the excitation of fs laser pulses \cite{Khorsand2013} as well as for pure elements \cite{Wietstruk2011}. For both materials, a transient ferromagnetic-like state is initiated. However, the latter is followed by magnetization recovery in the case of Tb-based alloy, while it is followed by magnetization reversal in the case of Gd-based alloy. These findings were attributed to the large difference between the spin-orbit coupling of Tb and Gd, and are a hint that switching mechanisms in Gd-based alloys and other ferrimagnetic alloys are distinct.

We now elucidate the switching process in Tb\textunderscript{27}Co\textunderscript{73} alloy films via the same set of experiments performed on GdFeCo and Pt/Co/Pt films. First, the response of the Tb\textunderscript{27}Co\textunderscript{73} films under the excitation of two consecutive single 35-fs pulses with a fluence of 19 mJ/cm$^2$ was probed by optical Faraday microscopy. This experiment showed results similar to Pt/Co/Pt films, as each of the two pulses induces thermal demagnetization of the irradiated area independently of the pulse polarization and the initial saturation as shown in Supplementary Fig. 3. Second, the magnetization change of a Tb\textunderscript{27}Co\textunderscript{73} based Hall cross under the action of 600 consecutive 35-fs pulses for different polarizations and initial saturations was electrically probed. As depicted in Fig. 5a,b, the response of the anomalous Hall voltage to ultrafast laser pulses is very similar to the one measured on Pt/Co/Pt Hall crosses, and is also governed by a two regimes process. The first part of the process corresponds to a step-like and heat-only demagnetization within the first 1 ms, as shown in the insets of Fig. 5a,b. Only after the demagnetization, the second part of the process occurs and consists in a helicity-dependent as well as gradual remagnetization taking place within a 100 ms timescale, indicating that 500 pulses are needed to reach a reliable switching. Thus, these experiments undoubtedly demonstrate that the AOS mechanism in TbCo alloys is similar to one of Pt/Co/Pt films. Note that in this Tb\textunderscript{27}Co\textunderscript{73} Hall cross, left-($\sigma$-) (resp. right-($\sigma$+)) circular polarized pulses switch the magnetization to down (resp. to up). However, as can be seen from Fig. 4, left-($\sigma$-) (resp. right-($\sigma$+)) circular polarized pulses switch the magnetization of the Pt/Co/Pt Hall cross to up (resp. to down). This finding is in agreement with previous studies and is attributed to the fact that the helicity of switching depends on the orientation of the Co sublattice magnetization and not on the direction of the net magnetization of the Tb dominated Tb\textunderscript{27}Co\textunderscript{73} \cite{Hassdenteufel2014}.
\\

\begin{figure}[h]
\begin{center}
\scalebox{1}{\includegraphics[width=7 cm, clip]{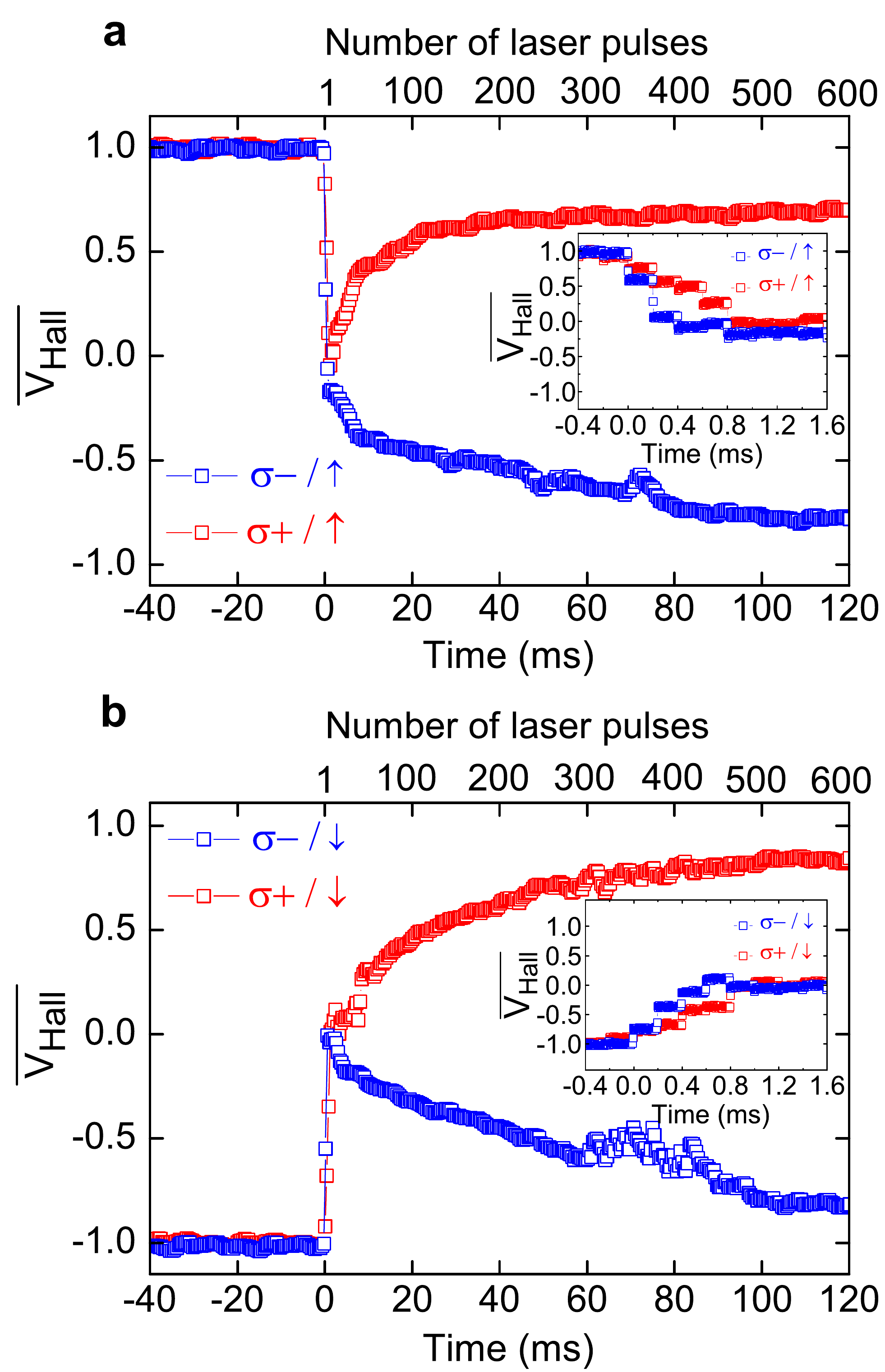}}
\end{center}
\makeatletter 
\renewcommand{\thefigure}{\textbf{5}}
\caption{\label{sample} \textbf{Multiple pulse helicity-dependent switching of TbCo Hall crosses}. (\textbf{a}) and (\textbf{b}) Helicity-dependent reversal of the anomalous Hall voltage of a Tb\textunderscript{27}Co\textunderscript{73} based Hall cross initially saturated ``up" and ``down" within a 120 ms timescale. The experiment is performed using a 35-fs laser beam with a 5 kHz repetition rate and a fluence of 19 mJ/cm$^2$. The corresponding number of laser pulses is shown in the upper row. The inset in (\textbf{a}) (resp. (\textbf{b})) shows the helicity-independent demagnetization of the Hall cross with an initial saturation magnetization ``up" (resp. ``down") occurring within the first 1 ms. }
\end{figure}

\noindent
\textbf{Discussion} 
\\
The main conclusion of this article is to distinguish between two different all-optical switching mechanisms. This was achieved by verifying the single-pulse helicity-independent switching of GdFeCo alloy films, and on the other hand by presenting a two regimes and cumulative switching of Pt/Co multilayers and TbCo alloy films. In regards to GdFeCo alloy films, the helicity-independent switching has already been presented in previous studies. It is attributed to a heat-only process, which is due to the distinct dynamics of the Gd and Fe sublattices of the ferrimagnet. Indeed, time-resolved XMCD measurements have shown that under the action a single fs pulse, Fe and Gd moments flip at two different timescales (300 fs for Fe spin and 1.5 ps for Gd spin), thus leading a transient alignment of the two moments despite of their antiferromagnetic coupling at the ground state \cite{Radu2011}. Note that this heat-only switching is achieved with any polarization at high fluences. Consequently, it was also demonstrated that GdFeCo films exhibit all-optical switching with only one circular helicity of light only for a narrow window of fluence, which is estimated to 1.5$\%$ of the threshold fluence for switching. This helicity-dependent switching of GdFeCo films was explained quantitatively with magnetic circular dichroism (MCD) \cite{Khorsand2012}. 

The experimental results presented in this paper for GdFeCo alloy films are fully consistent with the previous data and the model proposed above. However, it cannot explain the switching mechanism for Pt/Co/Pt and TbCo films. For these materials, the most intriguing question that remains to be answered is the microscopic origin of the magnetization evolution in the Hall crosses, which consists in a step-like helicity-independent demagnetization followed by a helicity-dependent remagnetization. Firstly, one important question is the fluence dependence of this reversal mechanism. For both Pt/Co/Pt and TbCo films, we additionally investigated the fluence dependence with a single pulse excitation. Only a thermal demagnetization is measured by varying the fluence from the switching threshold of the sweeping beam measurements to the damage fluence threshold. We have also verified the fluence dependence with multiple pulse illumination using the AHE characterization. The step-like demagnetization is maintained for a large range of fluence which is estimated to 20$\%$ of the threshold fluence for switching. However, the remagnetization vanishes for low fluences due to the spatial shift of the all-optical switching rim as shown in Supplementary Fig. 4. These findings indicate that the laser fluence does not play a primary role in the cumulative and two-step switching of Pt/Co/Pt and TbCo alloy films. Furthermore, we attribute the helicity-independent demagnetization to a heat-only induced reduction of the demagnetizing energy, leading to a demagnetized state. Such demagnetized state consists of magnetic domains with random orientations up and down, whose size is smaller than the Hall cross area. To prove that this domain formation is responsible for the step-like drops of the Hall voltage and to exclude that a gradual and persistent heating of the cross is responsible for this behavior, we added a small external magnetic field of approximatively 20 mT. In this case, the Hall voltage and thus also the magnetization of the laser-illuminated Hall cross remained unchanged, which excludes a significant heating on time scales longer than 1 $\mu$s (see Supplementary Fig. 2). Further, the demagnetization of the studied 5-$\mu$m-wide Hall crosses follows a step-like drop induced by the first 6 pulses. By measuring a similar step-like demagnetization in a 20-$\mu$m-wide Pt/Co/Pt Hall cross using the same laser fluence, we have found that 30 pulses are needed to reach the demagnetized state, thus indicating that the number of pulses required to obtain the complete demagnetization is related to the Hall cross size (see Supplementary Fig. 5).

The more intriguing part of this observed magnetization mechanism is the helicity-dependent remagnetization that takes place only after the helicity-independent demagnetization. A potential scenario of this remagnetization would be a helicity-dependent domain wall motion. Indeed, once the heat-only induced multidomain state is obtained, a space-dependent symmetry breaking might arise from a different absorption of the two circular helicities in the different domains. This particular point is in agreement with the previous studies on GdFeCo films highlighting the role of magnetic circular dichroism (MCD) in the all-optical switching process \cite{Khorsand2012}, even though the final result of the switching is quite different. A domain-dependent and thus space-dependent absorption would create a non-uniform heat distribution in the multi-domain state, thus stimulating spin currents via the spin Seebeck effect (SSE) \cite{Uchida2008,Adachi2013}. Following this, the spin currents passing through a domain wall might induce a spin transfer torque (STT) leading to a helicity-dependent shift of domain walls and thus to a gradual remagnetization \cite{Tatara2004,Hinzke2011,Ryu2013}. Note that time-dependent thermal gradient might induce a pure thermal motion of magnetic domain walls \cite{Torrejon2012}. Moreover, superdiffusive spin currents might also play a major role in this process, as it has already been shown in a previous study on GdFeCo films \cite{Graves2013}. One last issue we discuss here is the reason why the process is not starting again after the magnetization reaches the steady state (see Figs 4-5), especially when the helicity which is used does not lead to switching. Thus, the question is if there is a fundamental difference between the initial and final state in this case. Please note in this context that the AHE is only probing the central area of the cross, thus a changed domain configuration in the surroundings of the probed area cannot be detected and might be the underlying difference to the initial state. Indeed, as already mentioned, all-optical switching in such materials occurs only in a localized rim at the edge of a demagnetized area \cite{Lambert2014}, which is in our case placed on the central area of the cross. This explanation is consistent with a remagnetization governed by domain wall motion-like mechanism.

In conclusion, we have experimentally investigated all-optical switching for continuous films and Hall crosses on three different materials, namely two ferrimagnetic GdFeCo and TbCo alloys and one ferromagnetic Pt/Co/Pt multilayer. We distinguished between two different all-optical switching mechanisms. The first mechanism is the single pulse and ultrafast switching as originally observed in ferrimagnetic GdFeCo alloys, whereas the second process measured for Pt/Co/Pt and TbCo films is a cumulative and two regimes swichting mechanism. More importantly, the newly discovered two regimes process consists of a helicity-independent (and step-like) demagnetization followed by a helicity-dependent gradual remagnetization. These findings provide intriguing new insights into the rich physics underlying the all-optical switching. \\

\noindent
\textbf{Methods} 
\\
\textbf{Samples preparation and microstructuring.} All samples were grown by DC magnetron sputtering. Alloys were prepared by co-sputtering and multilayers by sequential deposition of layers. The direction of the easy axis, the saturation magnetization and the coercive field were determined using SQUID magnetometry. Samples were patterned into 5-$\mu$m-wide Hall crosses using UV lithography with Ar\textsuperscript{+} ion etching down to the glass substrate. Thereafter, metal contacts were obtained via lift-off technique after deposition of Ta(10 nm)/Pt(100 nm) by DC magnetron sputtering. \\

\noindent
\textbf{All-optical switching measurements.} To perform optical excitation, we use a Ti: sapphire fs-laser with a 5-kHz repetition rate, a wavelength of 800 nm (1.55 eV), and pulse duration of 35 fs. The Gaussian beam spot is focused with a FWHM of approximately 60 $\mu$m. A quarter-wave plate is used to transform the linear polarized light ($\pi$) into right- ($\sigma$+) and left- ($\sigma$-) handed circularly polarized light, whereas a half-wave plate is used to adjust the laser power. Single pulses are selected using a pulse-picker. The samples are excited from their bottom side with typical laser fluences of 10-14 mJ/cm$^2$ for ferromagnetic multilayers and 19 mJ/cm$^2$ for ferrimagnetic alloys. 

Prior to being optically excited, continuous films are saturated with an external magnetic field, whereas no magnetic field is applied during the optical excitation. The response of the magnetic films under the action of single pulses was studied using a Faraday microscope to image the magnetic domains. In patterned Hall crosses, the fs laser beam is maintained at a fixed position where the center of the beam is off-centered to overlap the beam's rim where AOS takes place with the central area of the Hall cross, as shown in Fig. 1b. Before the illumination, the Hall cross is saturated under an external magnetic field applied along the z axis aiming to quantify the anomalous Hall voltage \textit{$V_{Hall}$}, which is proportional to the average value of \textbf{\textit{M}}\textbf{\textit{$\hat{e}_\mathrm{z}$}} at the Hall cross central region \cite{Luttinger1958,Berger1970}, for the two saturated magnetic states \textbf{\textit{M}}\textbf{\textit{$\hat{e}_\mathrm{z}$}} or -\textbf{\textit{M}}\textbf{\textit{$\hat{e}_\mathrm{z}$}}. During the illumination, no external magnetic field is applied and \textit{$V_{Hall}$} is measured continuously, thus indicating the magnetic state of the Hall cross under fs laser excitations. The typical value of applied current is 1 mA and multiple pulse induced magnetization change is quantified by measuring the anomalous voltage Hall with a temporal resolution of 1 $\mu$s. The anomalous Hall voltage presented in Figures 2,4,5 is normalized to its saturation value. All measurements are performed at room temperature.

\noindent
\textbf{Acknowledgments} 

\noindent
We would like to thank G. Lengaigne for technical assistance with lithography process, as well P. Riego and P. Vallobra for technical assistance with optical measurements. We would also like to thank E.E. Fullerton, V. Lomakin, Y. Fainman and R. Medapalli for fruitful discussions. This work was supported by the ANR-NSF Project, ANR-13-IS04-0008-01, ``COMAG" by the ANR-Labcom Project LSTNM, ANR-15-CE24-0009 UMANI  and by the Universit\'{e} de la Grande Region (UniGR funded P. Pirro Post-Doc). Experiments were performed using equipment from the TUBE - Daum funded by FEDER (EU), ANR, the Region Lorraine and Grand Nancy. \\

\noindent
\textbf{Author contributions}  \\
\noindent
G.M., M.H. and S.M. designed and coordinated the project; M.S.E.H., C.-H.L. and M.H. grew, characterized and optimized the samples. M.S.E.H., P.P., S.P.-W. and F.M. patterned, optimized and measured the Hall cross devices. M.S.E.H., P.P., C.-H.L., Y.Q. and G.M. operated the Faraday microscope and the pump laser set-up. M.S.E.H. and S.M. coordinated work on the paper with contributions from P.P., S.P.-W., M.H., F.M., G.M. and regular discussions with all authors. \\

\noindent
\textbf{Additional information} 

\noindent
The authors declare no competing financial interests.
\newpage

\renewcommand{\figurename}{\textbf{Supplementary Figure}}

\preprint{}
\maketitle

\center{\Large{\textbf{Supplementary Information}}}
\vspace{1 cm}
\center{\large{\textbf{Supplementary Figure 1}}}
\vspace{1 cm}

\begin{figure}[h]
\begin{center}
\scalebox{1}{\includegraphics[width=12 cm, clip]{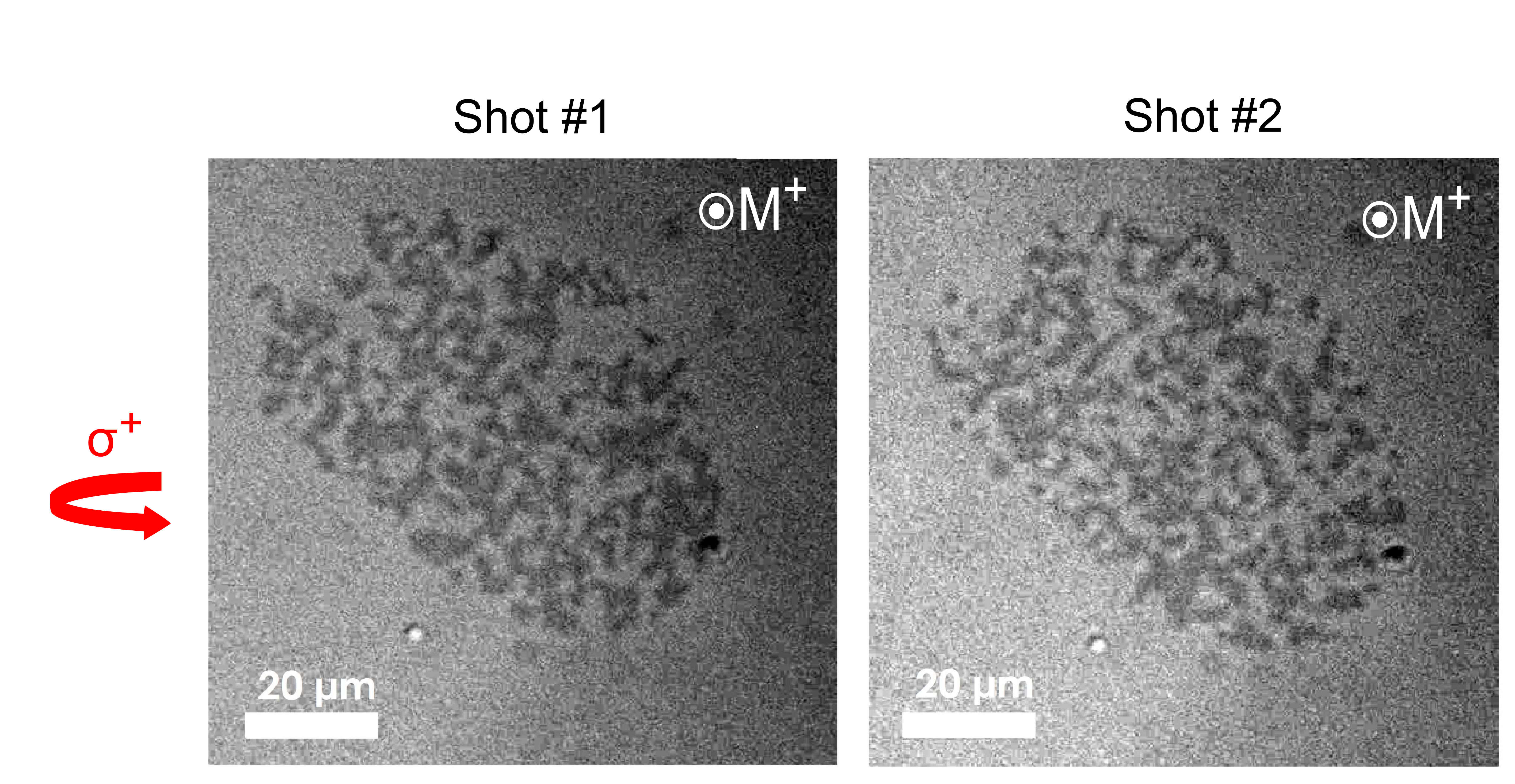}}
\end{center}
\setcounter{figure}{0}
\makeatletter 
\renewcommand{\thefigure}{\textbf{1}}
\caption{\label{Supplementary Figure 1}\textbf{Single shot induced thermal demagnetization of Pt/Co/Pt continuous films.} Magneto-optical Faraday images of a Pt(4.5 nm)/Co(0.6 nm)/Pt(4.5 nm) continuous film with initial magnetization saturation ``up", illuminated with two consecutive right-circularly polarized pulses with an energy per pulse of 14 mJ/cm$^2$. The dark contrast corresponds to a reversal to down. Each of the two laser pulses induces thermal demagnetization of the irradiated area and the rim of helicity-dependent switching does not emerge.}
\end{figure}

\newpage
\center{\large{\textbf{Supplementary Figure 2}}}
\vspace{1 cm}

\begin{figure}[h]
\begin{center}
\scalebox{1}{\includegraphics[width=16 cm, clip]{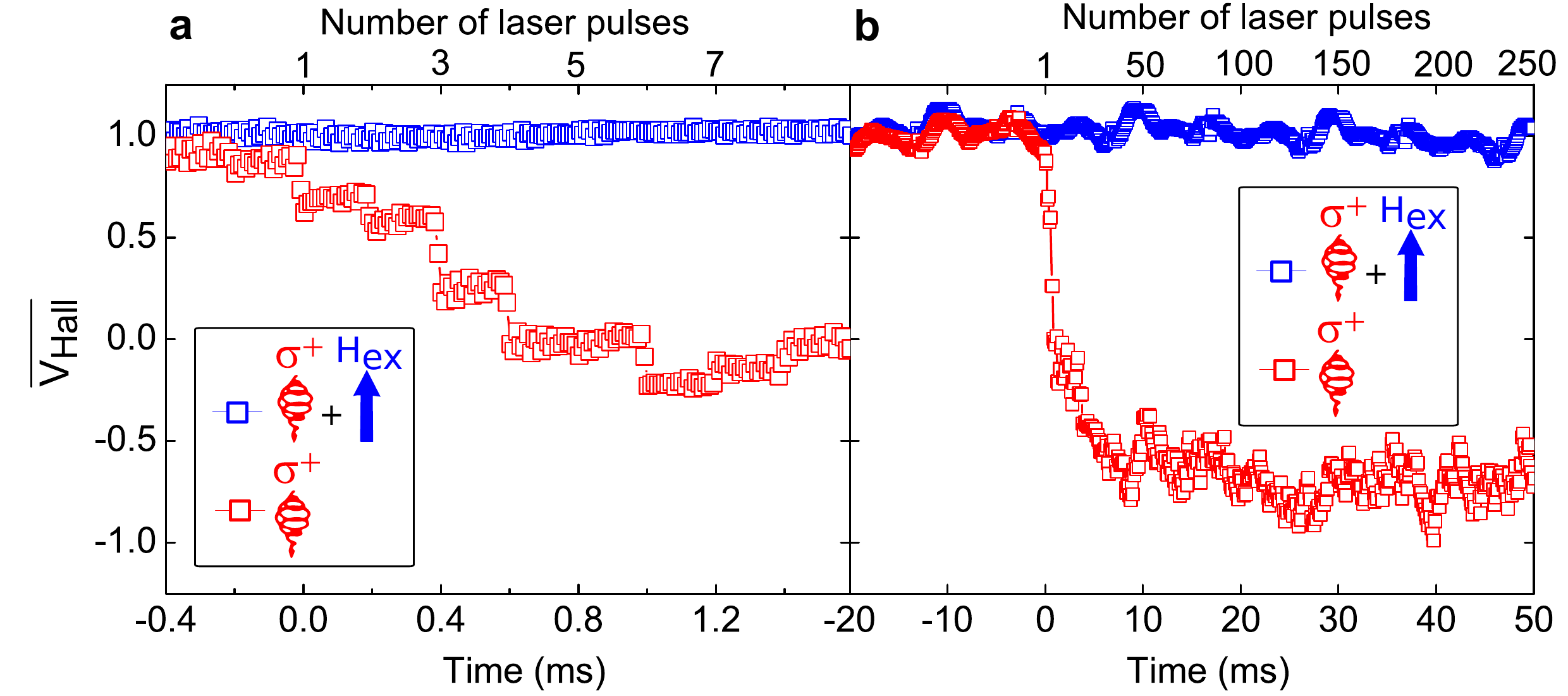}}
\end{center}
\makeatletter 
\renewcommand{\thefigure}{\textbf{2}}
\caption{\textbf{\label{sample}}\textbf{Saturation with a small external magnetic field of multiple pulse-illuminated Pt/Co/Pt Hall crosses.} (\textbf{a}) and (\textbf{b}) Time evolution of the anomalous Hall voltage of a 5-$\mu$m-wide Pt(4.5 nm)/Co(0.6 nm)/Pt(4.5 nm) Hall cross initially saturated ``up", under the action of a 35-fs laser beam with a 5 kHz repetition rate and a laser fluence of 10 mJ/cm$^2$, with (blue curve) and without (red curve) adding a small external magnetic field \textit{$H_{ex}$} of approximatively 20 mT. The corresponding number of laser pulses is shown in the upper row.} 
\end{figure}

\newpage
\center{\large{\textbf{Supplementary Figure 3}}}
\vspace{1 cm}

\begin{figure}[h]
\begin{center}
\scalebox{1}{\includegraphics[width=12 cm, clip]{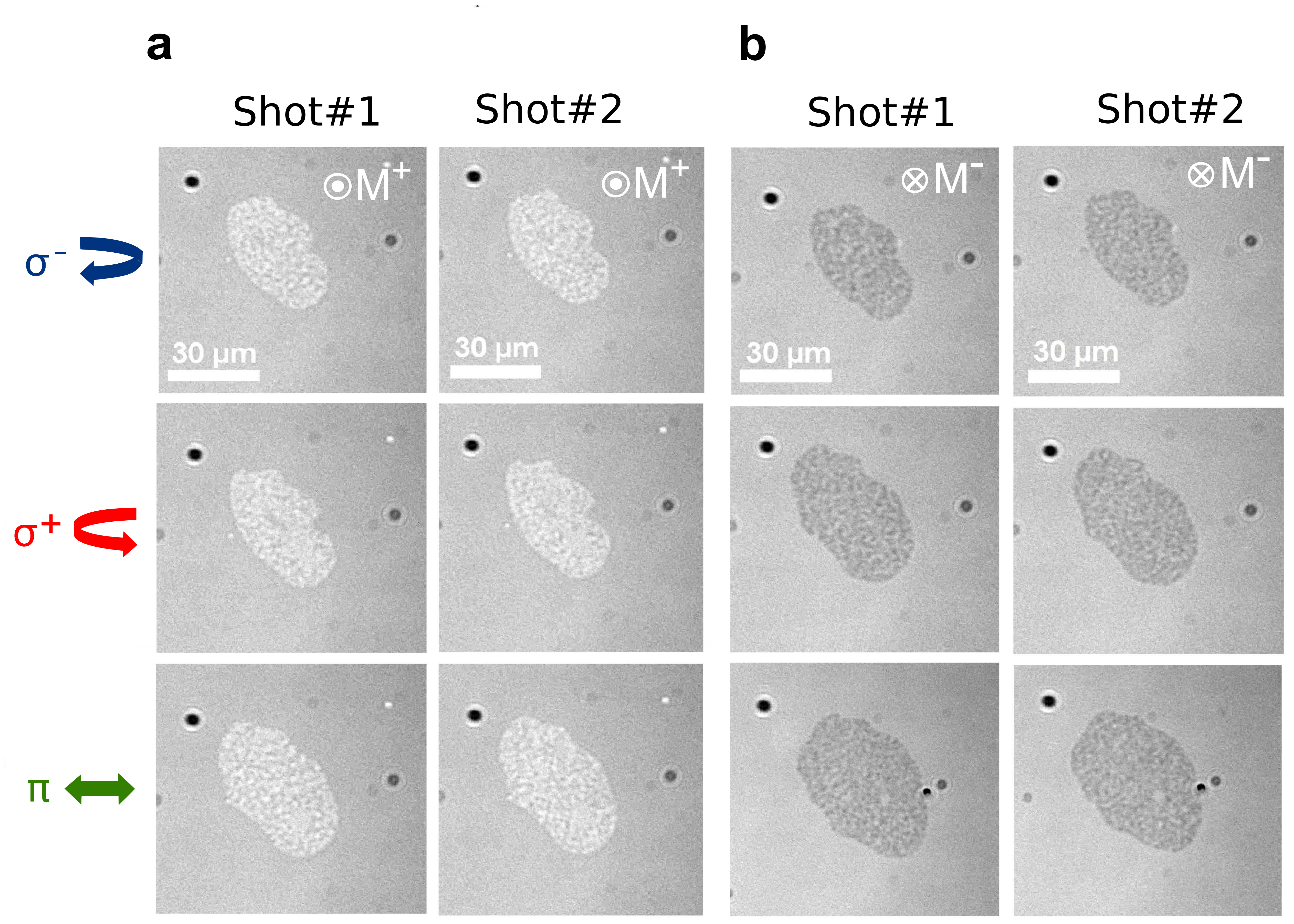}}
\end{center}
\makeatletter 
\renewcommand{\thefigure}{\textbf{3}}
\caption{\label{sample}\textbf{Single shot induced thermal demagnetization of TbCo continuous films.} (\textbf{a}) and (\textbf{b})  Magneto-optical Faraday images of a TbCo alloy film, with initial magnetization saturation ``up" and ``down", illuminated with two consecutive pulses with three different polarizations and with an energy per pulse of 19 mJ/cm$^2$. Each of the two laser pulses induces thermal demagnetization of the irradiated area. The white (resp. dark) contrast in (\textbf{a})  (resp. (\textbf{b})) corresponds to a reversal to down (resp. to up).}
\end{figure}

\newpage
\center{\large{\textbf{Supplementary Figure 4}}}
\vspace{1 cm}

\begin{figure}[h]
\begin{center}
\scalebox{1}{\includegraphics[width=16 cm, clip]{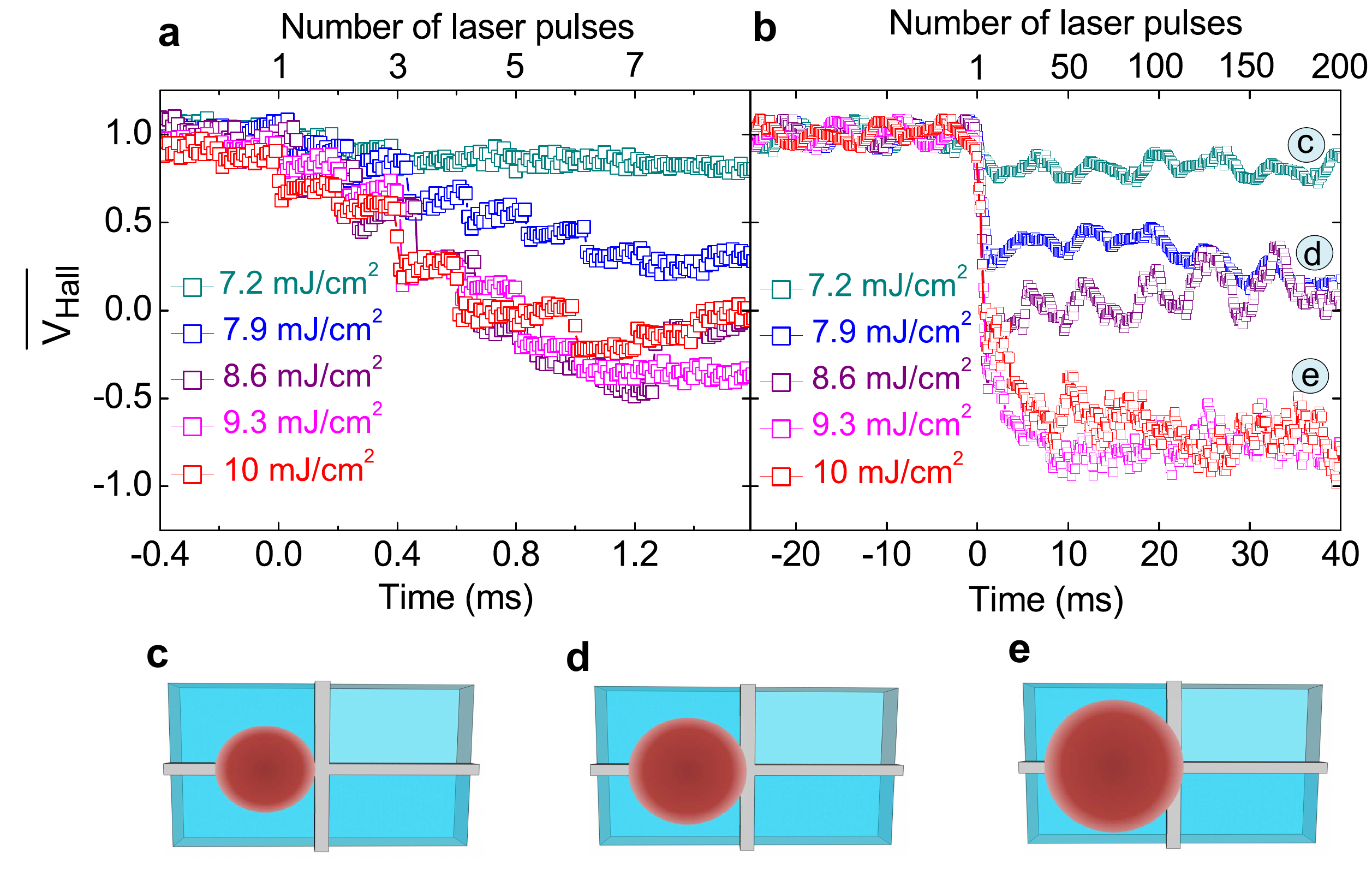}}
\end{center}
\makeatletter 
\renewcommand{\thefigure}{\textbf{4}}
\caption{\label{sample}\textbf{Fluence dependence of the helicity-dependent switching of 5-$\boldsymbol\mu$m-wide Pt/Co/Pt Hall crosses.} (\textbf{a}) and (\textbf{b}) Time evolution of the anomalous Hall voltage of a 5-$\mu$m-wide Pt(4.5 nm)/Co(0.6 nm)/Pt(4.5 nm) Hall cross initially saturated ``up", under the action of a 35-fs right-circularly polarized laser beam with a 5 kHz repetition rate and a laser fluence ranging from 7.2 mJ/cm$^2$ to 10 mJ/cm$^2$. The corresponding number of laser pulses is shown in the upper row. (\textbf{c}), (\textbf{d}) and (\textbf{e}), schematic evolution of the diameter of the helicity-dependent switching rim for laser fluences of 7.2 mJ/cm$^2$, 7.9-8.6 mJ/cm$^2$ and 9.3-10 mJ/cm$^2$, respectively.}
\end{figure}

\newpage
\center{\large{\textbf{Supplementary Figure 5}}}
\vspace{1 cm}

\begin{figure}[h]
\begin{center}
\scalebox{1}{\includegraphics[width=12 cm, clip]{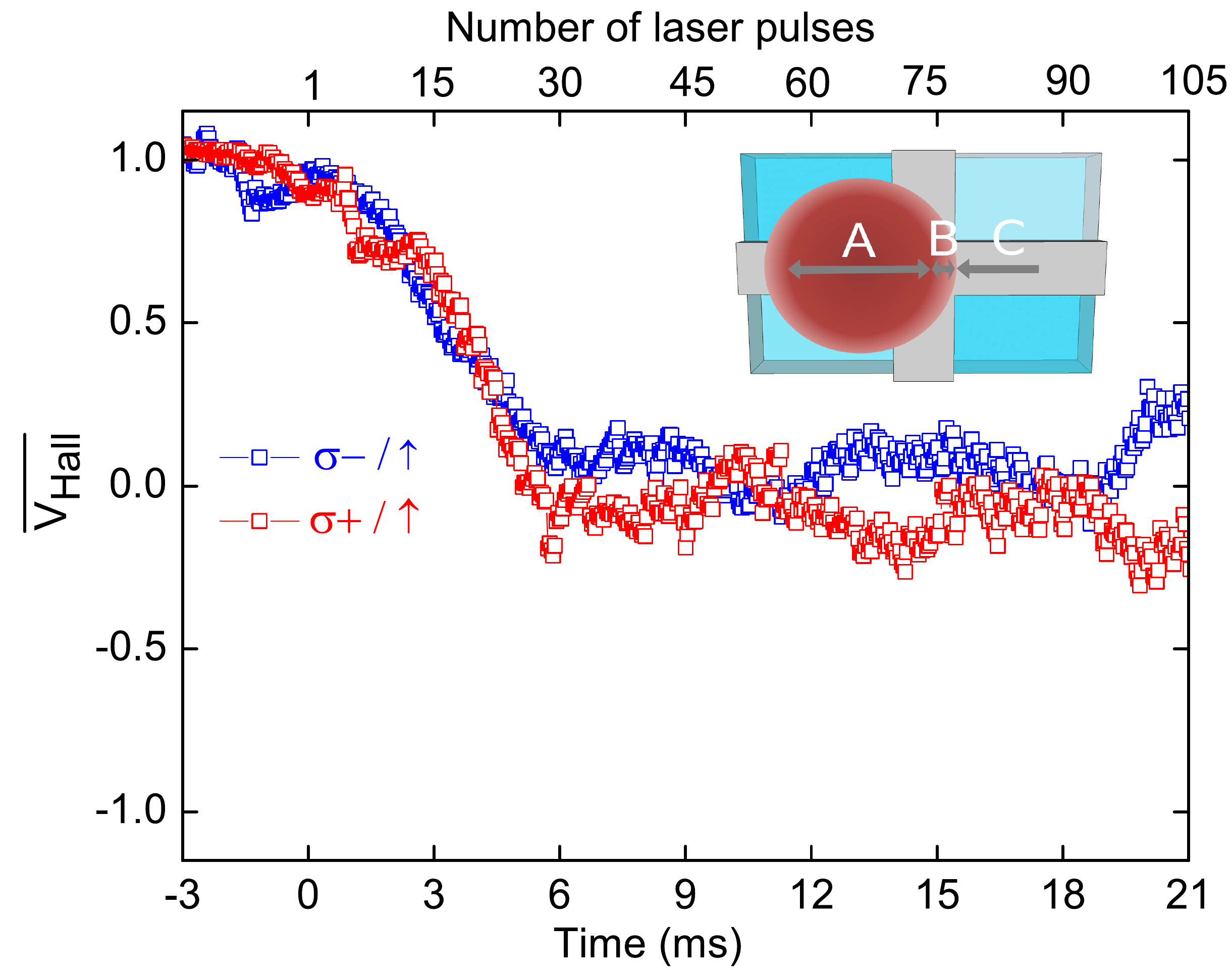}}
\end{center}
\makeatletter 
\renewcommand{\thefigure}{\textbf{5}}
\caption{\label{sample}\textbf{Multiple pulse thermal demagnetization of 20-$\boldsymbol\mu$m-wide Pt/Co/Pt Hall crosses.} Time evolution of the anomalous Hall voltage of a 20-$\mu$m-wide Pt(4.5 nm)/Co(0.6 nm)/Pt(4.5 nm) Hall cross initially saturated ``up", under the action of a 35-fs laser beam with a 5 kHz repetition rate and a laser fluence of 10 mJ/cm$^2$. The corresponding number of laser pulses is shown in the upper row. A helicity-independent demagnetization occurs within the first 6 ms (30 pulses) followed by an oscillation of the Hall voltage around the zero value. The inset shows the schematic representation of the different areas of the fs laser beam: ``A" where multiple magnetic domains are obtained, ``B" the helicity-dependent switching rim; and ``C" where no change of magnetization is induced. The remagnetization does not take place since the helicity-dependent switching rim (2-4 $\mu$m size) only overlaps with a small area of the 20-$\mu$m-wide Hall cross. }
\end{figure}
\end{document}